\documentclass[letterpaper,10pt,twocolumn]{article}
\usepackage{usenix}
\usepackage{times}
\usepackage{amsmath, dsfont,latexsym,color,amssymb,setspace,amsthm}
\usepackage{url}
\usepackage{xspace}
\usepackage[]{grffile}
\usepackage{color}
\usepackage{authblk}
\usepackage{relsize}
\usepackage{algorithm}
\usepackage{algorithmicx}
\usepackage{multicol}
\usepackage[noend]{algpseudocode}
\usepackage{tikz}
\usepackage{amsmath}
\usepackage{amssymb}
\usepackage{datetime}
\usepackage{inconsolata}
\usepackage[T1]{fontenc}
\usepackage{paralist}
\usepackage{titlesec}
\usepackage{breakurl} 
\usepackage[breaklinks]{hyperref}
\usepackage{enumitem}

\definecolor{darkred}{rgb}{0.8,0,0}

\hypersetup{
    colorlinks = true,
    linkcolor = darkred,
    citecolor = {darkred},
    urlcolor = {darkred}
}

\titlespacing{\paragraph}{%
  0pt}{
  0.5\baselineskip}{
  1em}

\titlespacing\section{0pt}{9pt plus 2pt minus 2pt}{5pt plus 2pt minus 2pt}
\titlespacing\subsection{0pt}{9pt plus 2pt minus 2pt}{5pt plus 2pt minus 2pt}
\titlespacing\subsubsection{0pt}{9pt plus 2pt minus 2pt}{5pt plus 2pt minus 2pt}

\newcommand{\monitor}[1]{monitor}
\newcommand{\Monitor}[1]{Monitor}
\newcommand{\monitoree}[1]{monitoree}
\newcommand{\Monitoree}[1]{Monitoree}
\newcommand{\omitit}[1]{}
\newcommand{\subject}{subject}
\newcommand{\subjects}{subjects}
\newcommand{\observer}{observer}
\newcommand{\system}{Rapid}
\newcommand{\observers}{observers}

\newcommand{\remove}{\text{{\sc remove}}}
\newcommand{\join}{\text{{\sc join}}}

\newtheorem{Thm}{Theorem}
\newtheorem{Lem}[Thm]{Lemma}
\newtheorem{Cor}[Thm]{Corollary}

\newenvironment{Proof}{\medbreak
  \noindent {\bf Proof:~}}{\unskip\nobreak\hfill\hskip 2em \qed\par\medbreak}

\newcommand{\CDfull}{multi-process cut}
\newcommand{\CD}{CD}
\newcommand{\Viewfull}{view-change}
\newcommand{\View}{VC}

\date{}
\usepackage{titling}
\begin{document}

\title{Stable and Consistent Membership at Scale with Rapid\vspace{-2ex}}
\author{
{\rm Lalith Suresh$^{1}$, Dahlia Malkhi$^{1}$, Parikshit Gopalan$^{1}$, Ivan Porto Carreiro$^{3}$, Zeeshan Lokhandwala$^{2}$}\\
$^{1}$VMware Research Group, $^{2}$VMware, $^{3}$One Concern
}

\maketitle

\begin{abstract}

We present the design and evaluation of \system{}, a distributed membership service.
At \system{}'s core is a scheme for \emph{\CDfull{} detection} (\CD) that revolves around two
key insights: (i) it suspects a failure of a process only after alerts
arrive from multiple sources, and (ii) when a group of processes experience
problems, it detects failures of the entire group, rather than conclude
about each process individually.
Implementing these insights translates into a simple membership algorithm 
with low communication overhead.

We present evidence that our strategy suffices to drive unanimous detection
almost-everywhere, even when complex network conditions arise, such as   one-way
 reachability problems, firewall misconfigurations, and high packet loss.
Furthermore, we present both empirical evidence and analyses that proves that
the almost-everywhere detection happens with high probability.  To complete
the design, \system{} contains a leaderless consensus protocol that converts
\CDfull{} detections into a \Viewfull{} decision. The resulting membership
service works both in fully decentralized as well as logically centralized
modes.

We present an evaluation of \system{} in moderately scalable cloud settings. \system{}
bootstraps 2000 node clusters 2-5.8x faster than prevailing tools such as
Memberlist and ZooKeeper, remains stable in face of complex failure scenarios,
and provides strong consistency guarantees. It is easy to integrate \system{} into
existing distributed applications, of which we demonstrate two.

\end{abstract}

\section{Introduction}
\label{sec:introduction}

Large-scale distributed systems today need to be provisioned and resized
quickly according to changing demand.  Furthermore, at scale, failures are not
the exception but the norm~\cite{barroso:warehouse, tailatscale}.  This makes
membership management and failure detection a critical component of any
distributed system.

Our organization ships standalone products that we do not operate ourselves.
These products run in a wide range of enterprise data center environments. In
our experience, many failure scenarios are not always crash failures, but
commonly involve misconfigured firewalls, one-way connectivity
loss, flip-flops in reachability, and some-but-not-all packets being dropped
(in line with observations by \cite{deanPacketLoss, networkIsReliable,
Gill:2011:UNF:2018436.2018477, turner2012failure,
Guo:2015:PLS:2785956.2787496}). We find that existing membership solutions
struggle with these common failure scenarios, despite being able to cleanly
detect crash faults. In particular, existing tools take long to, or never
converge to, a stable state where the faulty processes are removed
(\S\ref{sec:motivation}).

We posit that despite several decades of research and production systems,
stability and consistency of existing membership maintenance technologies
remains a challenge. In this paper, we present the design and implementation
of \emph{\system{}}, a scalable, distributed  membership system that provides both
these properties. We discuss the need for these properties below, and
present a formal treatment of the service guarantees we require in
\S\ref{sec:design}.

\paragraph{Need for stability.}  Membership changes in distributed systems
trigger expensive recovery operations such as failovers and data migrations.
Unstable and flapping membership views therefore cause applications to
repeatedly trigger these recovery workflows, thereby severely degrading
performance and affecting service availability.   This was the case in several
production incidents reported in the Cassandra~\cite{cassandra6127,
cassandra3831} and  Consul~\cite{consul916, consul1212, consul1337} projects.
In an end-to-end experiment, we also observed a 32\% increase in throughput
when replacing a native system's failure detector with our solution that
improved stability (see \S\ref{sub:end_to_end_evaluation} for details).

Furthermore, failure recovery mechanisms may be faulty \emph{themselves} and
can cause catastrophic failures when they run
amok~\cite{hotosFailureRecovery, Gunawi:2016:WCS:2987550.2987583}. Failure
recovery workflows being triggered ad infinitum have led to
Amazon EC2 outages~\cite{awsOutage1, awsOutage2, awsOutage3}, Microsoft
Azure outages~\cite{azureServiceDiscovery, Huang:2017:GFA:3102980.3103005}, and
``killer bugs'' in Cassandra and
HBase~\cite{Gunawi:2014:BLC:2670979.2670986}.

Given these reasons, we seek to avoid frequent oscillations of the membership
view, which we achieve through stable failure detection.

\paragraph{Need for consistent membership views.}
Many systems require coordinated failure recovery, for example,
to correctly handle data re-balancing in storage systems \cite{autosharder,
Cowling:2009:CLM:1855807.1855819}. Consistent changes to the membership view
simplify reasoning about system behavior and the development of dynamic
reconfiguration mechanisms~\cite{Shraer:2012:DRP:2342821.2342860}.  

Conversely, it is challenging to build reliable clustered services on
top of a weakly consistent membership service~\cite{cassandra9667}.
Inconsistent view-changes may have detrimental effects.
For example, in sharded systems that rely on consistent hashing, an inconsistent
view of the cluster leads to clients directing requests to servers that do not
host the relevant keys \cite{cassandra11740, autosharder}. 
In Cassandra, the lack of consistent membership causes nodes to duplicate data
re-balancing efforts when concurrently adding nodes to a
cluster~\cite{cassandra9667} and also affects
correctness~\cite{cassandra11740}. To work around the lack of consistent
membership, Cassandra ensures that only a single node is joining the cluster
at any given point in time, and operators are advised to wait \emph{at least
two minutes} between adding each new node to a cluster~\cite{cassandra9667}.
 As a consequence,
bootstrapping a 100 node Cassandra cluster will take three hours and twenty
minutes, thereby significantly slowing down provisioning~\cite{cassandra9667}.

For these reasons, we seek to provide strict consistency, where membership
changes are driven by agreement among processes. 
Consistency adds a layer of safety above the failure detection layer and guarantees
the same membership view to all non-faulty processes.

\subsection*{Our approach}

\system{} is based on the following fundamental insights that bring stability
and consistency to \emph{both} decentralized and logically centralized membership
services:

\textbf{Expander-based monitoring edge overlay.} To scale monitoring load,
\system{} organizes a set of processes (a \emph{configuration}) into a stable
failure detection topology comprising \emph{\observers{}} that monitor and
disseminate reports about their communication \emph{edges} to their
\emph{\subjects{}}.  The monitoring relationships between processes forms a
directed expander graph with strong connectivity properties, which ensures with a high probability that healthy
processes detect failures.  We interpret multiple reports about a subject's
edges as a high-fidelity signal that the subject is faulty.

\textbf{Multi-process cut detection.} For stability, processes in \system{} (i) suspect a faulty
process $p$ only upon receiving alerts from multiple observers of $p$, and (ii)
delay acting on alerts about different processes until the churn
stabilizes, thereby converging to detect a global, possibly a multi-node
\emph{cut} of processes to add or remove from the membership. 
This filter is remarkably simple to implement, yet it suffices by itself to achieve
\emph{almost-everywhere agreement} -- unanimity among a large fraction of processes 
about the detected cut.

\textbf{Practical consensus.} For consistency, we show that converting almost-everywhere agreement 
into full agreement is practical even in 
large-scale settings. \system{}'s consensus protocol drives configuration changes by a
low-overhead, leaderless protocol in the common case: every process simply
validates consensus by counting the number of identical cut detections.
If there is a quorum containing three-quarters of the membership set 
with the same cut, then without a leader or
further communication, this is a safe consensus decision.

\system{} thereby ensures all participating processes see a strongly consistent
sequence of membership changes to the cluster, while ensuring that the system
is stable in the face of a diverse range of failure scenarios.

In summary, we make the following key contributions: 

$\bullet$ Through measurements, we demonstrate that prevailing membership
solutions guarantee neither stability nor consistency in the face of
complex failure scenarios.

$\bullet$ We present the design of \system{}, a scalable membership service that
is robust in the presence of diverse failure scenarios while providing strong
consistency. \system{} runs both as a decentralized as
well as a logically centralized membership service.

$\bullet$ In system evaluations, we demonstrate how \system{}, despite offering
much stronger guarantees, brings up 2000 node clusters 2-5.8x faster than
mature alternatives such as Memberlist and ZooKeeper. We demonstrate
\system{}'s robustness in the face of different failure scenarios such as
simultaneous node crashes, asymmetric network partitions and heavy packet
loss. \system{} achieves these goals at a similar cost to existing solutions.

$\bullet$ Lastly, we also report on our experience running \system{} to power two
applications; a distributed transactional data platform and service discovery of a fleet of web servers.

\section{Motivation and Related work}
\label{sec:related_work}

Membership solutions today fall into two categories. They are either managed for
a cluster through an auxiliary service \cite{zookeeper, hadoop}, or they are
gossip-based and fully decentralized \cite{consulLink, serf, cassandra, akka,
redis, eurosysOrleansOpt, uber_gossip, scylladb}.

We experimented with the behavior of  three widely adopted systems in the 
presence of
network failure scenarios: \emph{(i)} of the first category, 
ZooKeeper~\cite{zookeeper}, and of the second, \emph{(ii)}
Memberlist~\cite{Memberlist},   the membership library used by
Consul~\cite{consulLink} and Serf~\cite{serf} and \emph{(iii)} Akka
Cluster~\cite{akka} (see \S\ref{sec:evaluation} for the detailed setup).
For ZooKeeper
and Memberlist, we bootstrap a 1000 process cluster with stand-alone agents that
join and maintain membership using these solutions (for Akka Cluster, we use
400 processes because it began failing for cluster sizes beyond 500). We
then drop 80\% of packets for 1\% of processes, simulating high packet loss
scenarios described in the literature~\cite{networkIsReliable, deanPacketLoss} that we have
also observed in practice.

\begin{figure}[t]
\centering
\includegraphics[width=0.45\textwidth]{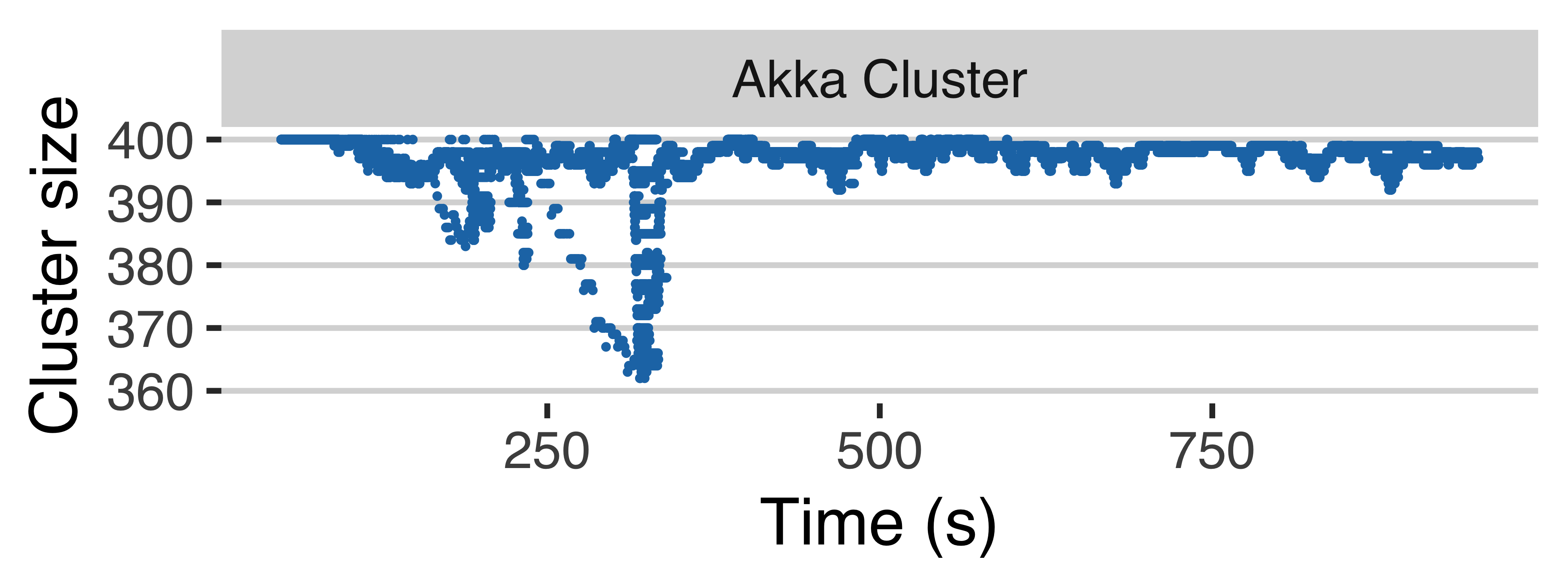}
\includegraphics[width=0.45\textwidth]{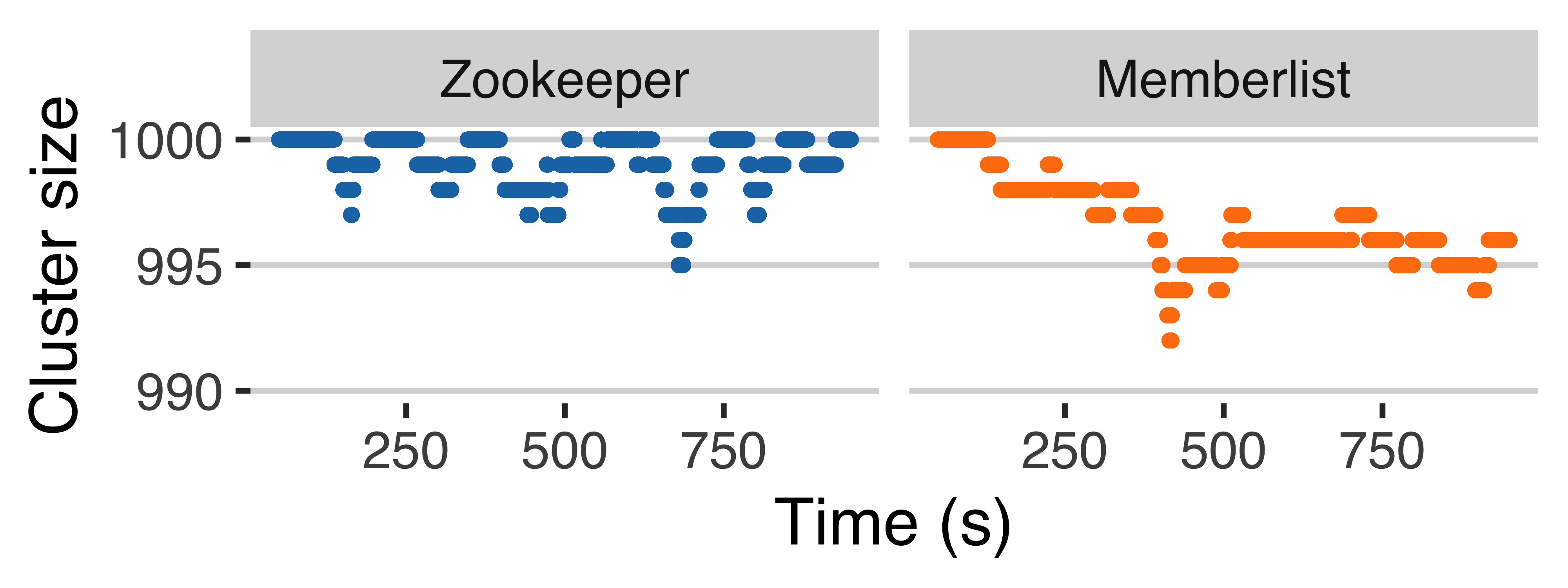}
\caption{Akka Cluster, ZooKeeper and Memberlist exhibit instabilities and inconsistencies when 1\% of processes experience 80\% packet loss (similar
to scenarios described in~\cite{networkIsReliable, deanPacketLoss}). 
Every process logs its own view of the cluster size every second, shown as one dot
along the time (X) axis.
Note, the y-axis range
does not start at 0.
X-axis points (or intervals) with different cluster size values represent inconsistent views
 among processes at that point (or during the interval).}
\label{fig:flipflopAcMlOnly}
\end{figure}

Figure~\ref{fig:flipflopAcMlOnly} shows a timeseries of membership sizes, as
viewed by each non-faulty process in the cluster (every dot indicates a single
measurement by a process). Akka Cluster is unstable as conflicting rumors
about processes propagate in the cluster concurrently, even resulting in
benign processes being removed from the membership. Memberlist and ZooKeeper
resist removal of the faulty processes from the membership set but are unstable over
a longer period of time. We also note extended periods of inconsistencies 
in the membership view.

Having found existing membership solutions to be unstable in the presence
of typical network faults, we now proceed to discuss the broader design
space.

\subsection{Comparison of existing solutions} \label{sec:motivation}

There are three membership service designs in use today, each of which provides
different degrees of resiliency and consistency.

\paragraph{Logically centralized configuration service.} 

A common approach to membership management in the industry is to
store the membership list in an auxiliary management service such as
ZooKeeper~\cite{zookeeper}, etcd~\cite{etcd}, or
Chubby~\cite{Burrows:2006:CLS:1298455.1298487}. 

The main advantage is that the membership logic remains simple: a few
processes maintain the ground truth of the membership list with strong
consistency semantics, and the remaining processes query this list
periodically.

The key shortcoming here is that reliance on a small cluster reduces the overall resiliency of the system:
connectivity issues to the cluster, or failures among the small set of cluster
members themselves, may render the service unavailable (this has led Netflix
to build solutions like Eureka~\cite{eureka,whynotzookeeper}). As the ZooKeeper
developers warn, this also opens up
new failure modes for applications that depend on an auxiliary service for
membership~\cite{zookeeperProgrammer}.

\paragraph{Gossip-based membership.}  van Renesse et
al.~\cite{vanRenesse:2009:GFD:1659232.1659238,
VanRenesse:2003:ARS:762483.762485} proposed managing membership by using
gossip to spread positive notifications (keepalives) between all processes.
If a process $p$ fails, other processes eventually remove $p$ after a timeout. 
Swim~\cite{das2002swim}  was proposed as a variant of that approach that
reduces the communication overhead;  it uses gossip to spread ``negative''
alerts, rather than regular positive notifications.

Gossip based membership schemes are widely adopted in a variety of
deployed systems. This includes Cassandra~\cite{cassandra}, Akka~\cite{akka},
ScyllaDB~\cite{scylladb}, Serf~\cite{serf}, Redis Cluster~\cite{redis},
Orleans~\cite{eurosysOrleansOpt}, Uber's Ringpop~\cite{uber_gossip}, Netflix's
Dynomite~\cite{dynomite}, and some systems at Twitter~\cite{twitter_gossip}.

The main advantage of gossip-based membership is resiliency  and graceful
degradation (they tolerate $N - 1$ failures). The key disadvantages include
their  weak consistency guarantees and the complex emergent behavior leading
to stability problems.

Stability is a key challenge in gossip-based membership: When communication
fails between two processes which are otherwise live and correct, there are
repeated accusations and refutations that may cause oscillations in the
membership views. As our investigation of leading gossip-based solutions
showed (Figure~\ref{fig:flipflopAcMlOnly}), these conflicting alerts lead to
complex emergent behavior, making it challenging to build reliable clustered
services on top of. Indeed, stability related issues with gossip are also
observed in production settings  (see, e.g., Consul~\cite{consul916,
consul1212, consul1337} and Cassandra~\cite{cassandra9667, cassandra11740,
cassandra6127}).

Lastly, FireFlies~\cite{Johansen:2015:FSS:2785582.2701418} is a decentralized
membership service that tolerates Byzantine members.  FireFlies organizes
monitoring responsibilities via a randomized $k$-ring topology  to provide a
robust overlay against Byzantine processes. While the motivation in FireFlies was
different, we believe it offers a solution for stability; accusations about a
process by a potentially Byzantine monitor are not acted upon until a
conservative, fixed delay elapses. If a process does not refute an accusation about
it within this delay, it is removed from the membership. However,
the FireFlies scheme is based on a gossip-style protocol involving accusations,
refutations, rankings, and disabling (where a process $p$ announces that a monitor
should not complain about it). As we show in upcoming
sections, our scheme is simple in comparison and requires little book-keeping per process. 
Unlike FireFlies, we aggregate
reports about a process $p$ from multiple sources to decide whether to
remove $p$, enabling coordinated membership changes with low overhead.

\paragraph{Group membership.}  
By themselves, gossip-based membership schemes do not address
consistency, and allow the membership views of processes to diverge.
In this sense, they may be considered more of failure detectors, than membership
services.

Maintaining membership with strict consistency guarantees
has been a focus in the field of fault tolerant state-machine
replication (SMR), starting with early foundations of
SMR~\cite{Lamport:1998:PP:279227.279229,
oki1988viewstamped,Schneider:1990:IFS:98163.98167}, and continuing with a
variety of group communication systems
(see~\cite{Chockler:2001:GCS:503112.503113} for a survey of GC works). In SMR
systems, membership is typically needed for selecting a unique primary and for
enabling dynamic service deployment.
Recent work on Census~\cite{Cowling:2009:CLM:1855807.1855819} scales dynamic
membership maintenance to a locality-aware hierarchy of domains. It provides
fault tolerance by running the view-change consensus protocol only among a sampled subset of
the membership set. 

These methods may be harnessed on top of a stable failure detection facility,
stability being orthogonal to the consistency they provide. As we show, our
solution uses an SMR technique that benefits from stable failure detection to
form fast, leaderless consensus.

\section{The \system{} Service}
\label{sec:design}

Our goal is to create a membership service based on techniques that apply
equally well to  both  decentralized as well as logically centralized designs.
For ease of presentation, we first describe the fully decentralized \system{}
service and its properties in this section, followed by its design in
\S\ref{sec:algorithm}.  We then relax the resiliency properties in
\S\ref{sec:algorithm_aux} for the logically centralized design.

\paragraph{API} Processes use the membership service by using the \system{} library
and invoking a call \textsc{join(host:port, seeds, view-change-callback)}.
Here, \textsc{host:port} is the process' TCP/IP listen address. Internally,
the join call assigns a unique logical identifier for the process (\textsc{id}). If a process departs from the
cluster either  due to a failure or by voluntarily leaving, it rejoins
with a new \textsc{id}.  This \textsc{id} is internal to \system{} and is not an
identifier of the application that is using \system{}. \textsc{seeds} is an initial 
set of process addresses known to everyone and
used to contact for bootstrapping. \textsc{view-change-callback} is used
to notify applications about membership change events.

\paragraph{Configurations} A configuration in \system{} comprises a configuration
identifier and a \emph{membership-set} (a list of processes).
Each process has a local \emph{view} of the configuration.  All processes use
the initial seed-list as a bootstrap configuration $C_0$. Every configuration change
decision triggers an invocation of the \textsc{view-change-callback} at all
processes, that informs processes about a new configuration and membership
set.

At time $t$, if $C$ is the configuration view of a majority of its members, we
say that  $C$ is the \emph{current configuration}. Initially, once a majority
of $C_0$ start, it becomes current.

\paragraph{Failure model} We assume that every pair of correct processes can
communicate with each other within a known transmission delay bound (an
assumption required for failure detection).
When this assumption is violated for a pair of (otherwise live) processes,
there is no obvious definition to determine which one of them is faulty
(though at least one is).  We resolve this using the parameters $L$ and $K$ as
follows. Every process $p$ (a \emph{subject}) is monitored by $K$
\emph{observer} processes. If $L$-of-$K$ correct observers cannot communicate
with a subject, then the subject is considered \emph{observably unresponsive}.
We consider a process faulty if it is either crashed or observably
unresponsive.

\paragraph{Cut Detection Guarantees}

Suppose that at time $t$, $C$ is the current configuration. Consider a subset
of processes $F \subset C$ where $\frac{|F|}{|C|} < \frac{1}{2}$. If all processes
in $C \setminus F$ remain non-faulty, then we guarantee that the \CDfull{}
will eventually be detected and a \Viewfull{} $C \setminus F$ installed
\footnote{The size of cuts $|F|$ we can detect is a function
of the monitoring topology. The proof appears in \S\ref{sec:guarantees_summary}.}
:

$\bullet$ Multi-process cut detection:
With high probability, every process in $C
\setminus F$ receives a \CDfull{} detection notification about $F$.  In \system{},
the probability is taken over all the random choices of the observer/subject
overlay topology, discussed in \S\ref{sec:algorithm}. The property we
use is that with high probability the underlying topology remains an
expander at all times, where the expansion is quantified in terms of
its second eigenvalue.

A similar guarantee holds for joins. If at time $t$ a set $J$ of
processes join the system and remain non-faulty, then every process in $C \cup
J$ is notified of $J$ joining.

Joins and removals can be combined: If a set of processes $F$ as above fails,
and a set of processes $J$ joins, then $(C \setminus F) \cup J$ is eventually
notified of the changes.

$\bullet$ View-Change:
Any \Viewfull{} notification in $C$ is by consensus, maintaining
\emph{Agreement} on the \Viewfull{} membership among
all correct processes in $C$; and \emph{Liveness}, provided a 
majority of $C$ (and of joiners $J$, if any) remain correct until the
\View{} configuration becomes current. 

Our requirements concerning configuration changes hold when the system has quiesced. During
periods of instability, intermediate detection(s) may succeed, but there is no
formal guarantee about them.

\paragraph{Hand-off}

Once a new configuration $C_{j+1}$ becomes current, we abstractly abandon
$C_j$ and start afresh:  New failures can happen within $C_{j+1}$ (for up to
a quarter of the membership set), and the Condition Detection and Consistent
Membership guarantees must hold.

We note that liveness of the consensus layer depends on a majority of both
$C_j$ and $C_{j+1}$ remaining correct to perform the `hand-off': Between the
time when $C_j$ becomes current and until $C_{j+1}$ does, no more
than a minority fail in either configuration. This dynamic model borrows
the dynamic-interplay framework of~\cite{gafni:hal-01206154,spiegelman-keidar-malkhi:disc2017}. %

\section{Decentralized Design}
\label{sec:algorithm}

\begin{figure}[t]
\centering
\includegraphics[width=0.47\textwidth, trim=10 300 0 0]{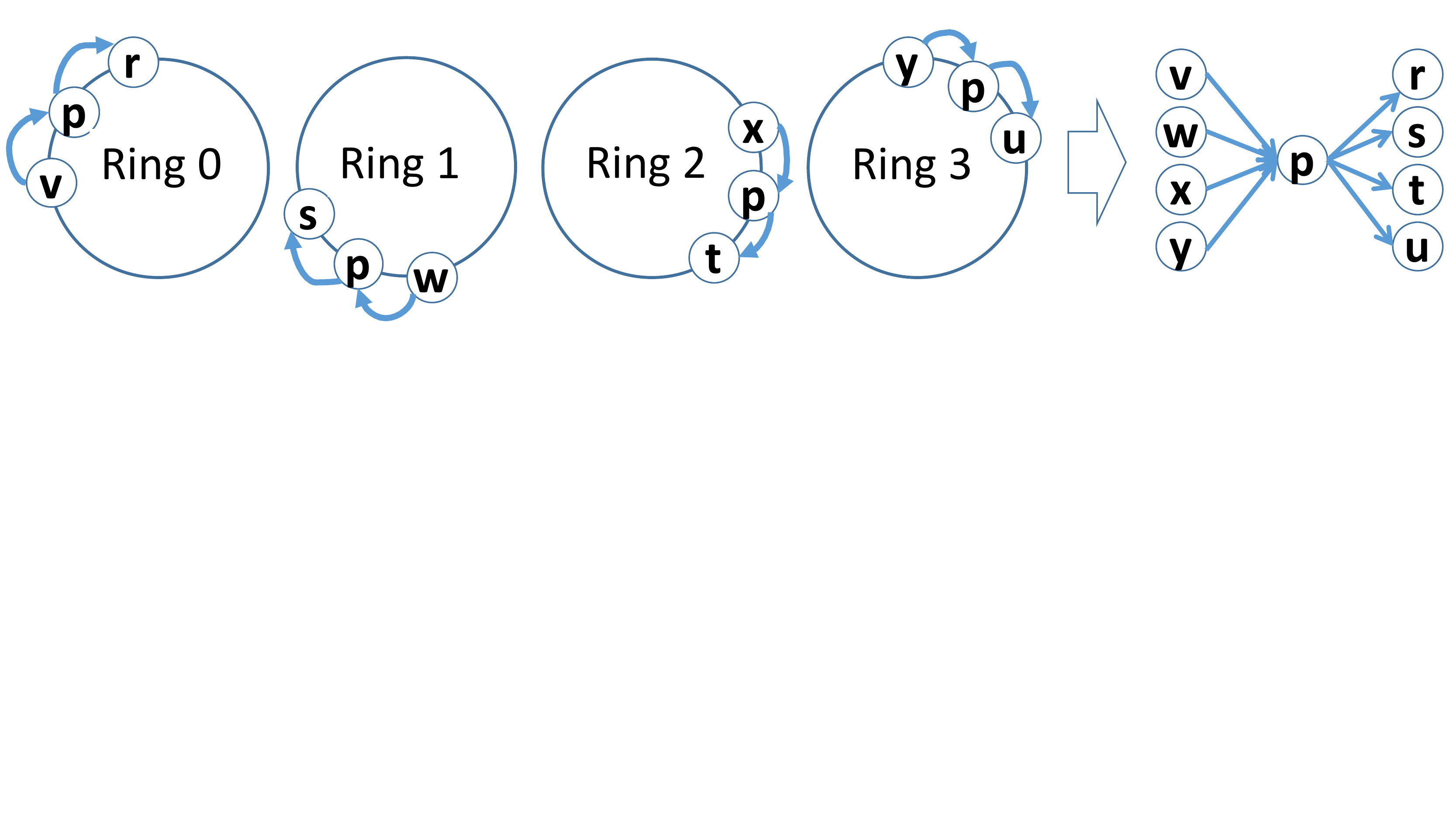}
\caption{$p$'s neighborhood in a $K=4$-Ring topology. $p$'s
\observers\
are $\{v, w, x, y\}$; $p$'s \subjects\ are $\{r, s, t, u\}$.}
\label{fig:monitoringTopology}
\end{figure}

\begin{figure}[t]
{\begin{center}
\includegraphics[width=0.5\textwidth,trim=0 350 0 0]{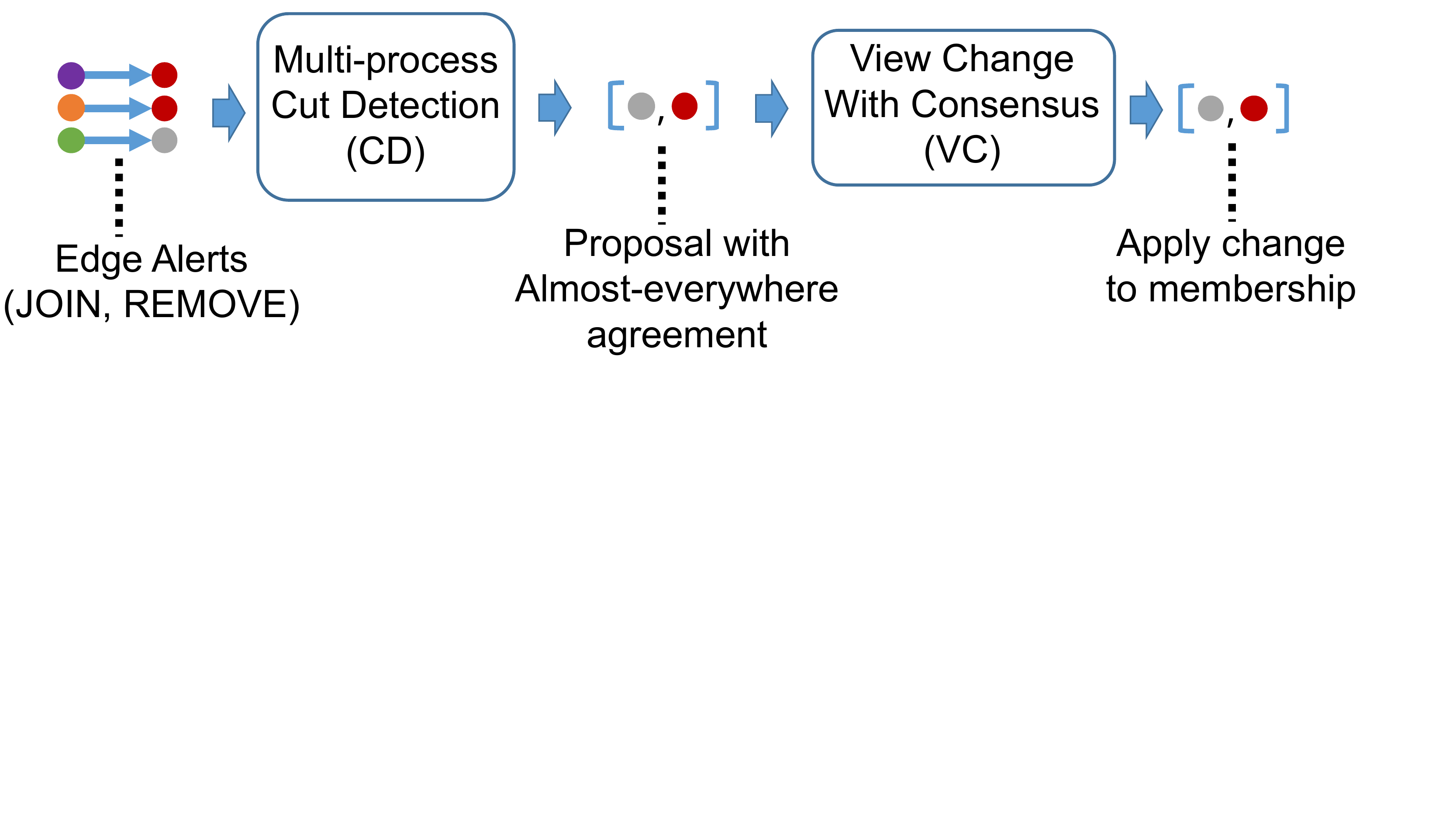}
\end{center}
}
\caption{Solution overview, showing the sequence of steps at each process 
for a single configuration change.}
\label{fig:solutionOverview}
\end{figure}

\system{} forms an immutable sequence of configurations driven through consensus
decisions. Each configuration may drive a single configuration-change
decision; the next configuration is logically a new system as in the virtual
synchrony approach~\cite{Birman:2010:HVS:2172338.2172344}. Here, we describe the algorithm for
changing a known current configuration $\mathcal{C}$, consisting of a
membership set (a list of process identities).  When clear from the context,
we omit $\mathcal{C}$ or explicit mentions of the configuration, as they are
fixed within one instance of the configuration-change algorithm.

We start with a brief overview of the algorithm, breaking it down to three components:
(1) a monitoring overlay;
(2) an almost-everywhere \CDfull\ detection (\CD); and
(3) a fast, leaderless \Viewfull\ (\View) consensus protocol.
The response to problems in \system{} evolves through these three components (see Figure~\ref{fig:solutionOverview}).

\paragraph{Monitoring} We organize processes into a monitoring topology such
that every process monitors $K$ peers and is monitored by $K$ peers.  A
process being monitored is referred to as a \emph{\subject{}} and a process
that is monitoring a subject is an \emph{\observer{}} (each process therefore
has $K$ \subject{}s and $K$ \observer{}s). The particular topology we employ
in \system{} is an \emph{expander graph}~\cite{basicFactsExpander} realized using
$K$ pseudo-random rings~\cite{secondEigenValueOfRRG}.  Other \observer/\subject\
arrangements may be plugged into our framework  without changing the rest of
the logic.

Importantly, this topology is deterministic over the membership set
$\mathcal{C}$; every process that receives a notification about a new
configuration locally determines its \subject{}s and  creates the required
monitoring channels.

There are two types of alerts generated by the monitoring component, \remove{}
and \join.  A \remove\mbox{} alert is broadcast by an \observer{} when there
are reachability problems to its \subject{}. A \join{} alert is broadcast by an
\observer{} when it is informed about a \subject{} joiner request. In this
way, both types of alerts are generated by multiple sources about the same
\subject.

\paragraph{Multi-process cut detection (\CD)} \remove\ and \join\ alerts are
handled at each process independently by a \CDfull{} detection (\CD)
mechanism. This mechanism collects evidence to support a single, stable multi-process 
configuration change proposal. It outputs the same cut
proposal \emph{almost-everywhere}; i.e., unanimity in the detection among a
large fraction of processes.

The \CD{} scheme with a $K$-degree monitoring topology has a constant
per-process per-round communication cost, and provides
stable \CDfull{} detection with almost-everywhere agreement.

\paragraph{View change (\View)} Finally, we use a consensus protocol that has
a fast path to agreement on a \Viewfull. If the protocol collects identical
\CD\ proposals from a \emph{Fast Paxos quorum} (three quarters) of the
membership, then it can decide in one step.   Otherwise, it falls back to
Paxos to form agreement on some proposal as a \Viewfull.

We note that other consensus solutions could use \CD{} as input and provide
\Viewfull{} consistency. \View{} has the benefit of a fast path to decision,
taking advantage of the identical inputs almost-everywhere.

We now present a detailed description of the system.

\subsection{Expander-based Monitoring}
\label{sub:edge_failure_detection_with_expander}

\system{} organizes processes into a monitoring topology that is an \emph{expander
graph}~\cite{basicFactsExpander}.
Specifically, we use the fact that a random K-regular graph is very
likely to be a good expander for $K \geq 3$~\cite{secondEigenValueOfRRG}. We construct $K$
pseudo-randomly generated rings with each ring containing the full list of
members.  A pair of processes ($o$, $s$) form an \observer/\subject\ \emph{edge}
if $o$ precedes $s$ in a ring. Duplicate edges are allowed and will have a
marginal effect on the behavior.
Figure~\ref{fig:monitoringTopology} depicts the neighborhood of a single process
$p$ in a $4$-Ring topology.

\paragraph{Topology properties.} Our monitoring topology has three key
properties. The first is \emph{expansion}:  the number of edges connecting two
sets of processes reflects the relative sizes of the set. This means
that if a small subset $F$ of  processes $V$ are faulty, we should see roughly
$\frac{\mid{V}\mid - \mid{F}\mid}{\mid{V}\mid}$   fraction of monitoring edges
to $F$ emanating from the set $V \setminus F$ of healthy processes (\S\ref{sec:guarantees_summary}). This ensures with high
probability that healthy processes detect failures, as long as the set
of failures is not too large. The size of failures we can detect depends on the expansion
of the topology as quantified by the value of its second eigenvalue.
Second, every process monitors $K$ \subject{}s, and is monitored by $K$ \observer{}s. Hence,
monitoring incurs $O(K)$ overhead per process per round, distributing the
load across the entire cluster. The fixed 
\observer/\subject{} approach distinguishes \system{} from gossip-based
techniques, supporting prolonged monitoring without sacrificing failure
detection scalability. At the same time, we compare well with the
overhead of gossip-based solutions (\S\ref{sec:evaluation}). Third, every process join or
removal results only in $2 \cdot K$ monitoring edges being added or
removed.

\paragraph{Joins}  New processes join by contacting a list of $K$ temporary
observers obtained from a seed process (deterministically assigned for each
joiner and $\mathcal{C}$ pair, until a configuration change reflects the join).
The temporary observers generate
independent alerts about joiners. In this way, multiple \join{} alerts are
generated from distinct sources, in a similar manner to alerts about failures.

\paragraph{Pluggable edge-monitor.} A monitoring edge between an observer and
its subject is a pluggable component in \system{}. With this design, \system{} can take
advantage of diverse failure detection and monitoring techniques, e.g.,
history-based adaptive techniques as used by popular frameworks like
Hystrix~\cite{hystrix} and Finagle~\cite{finagle}; phi-accrual failure
detectors~\cite{Defago04thephi}; eliciting indirect probes
~\cite{das2002swim};  flooding a suspicion and allowing a timeout period for
self-rebuttal~\cite{Johansen:2015:FSS:2785582.2701418}; using cross-layer
information~\cite{falcon}; application-specific health checks; and others.

\paragraph{Irrevocable Alerts}
\label{sub:disseminating_alerts}
When the edge-monitor of an \observer{} indicates an existing \subject{} is
non-responsive, the \observer{} broadcasts a \remove{} alert about the
\subject{}. Given the high fidelity made possible with our stable edge
monitoring, these alerts are considered \emph{irrevocable}, thus \system{}
prevents spreading conflicting reports. When contacted by a \subject, a
temporary \observer{}  broadcasts \join{} alert about the \subject.

\subsection{Multi-process Cut Detection}
\label{sec:almosteverywhere}

\begin{figure}[t]
\centering
\includegraphics[width=0.23\textwidth]{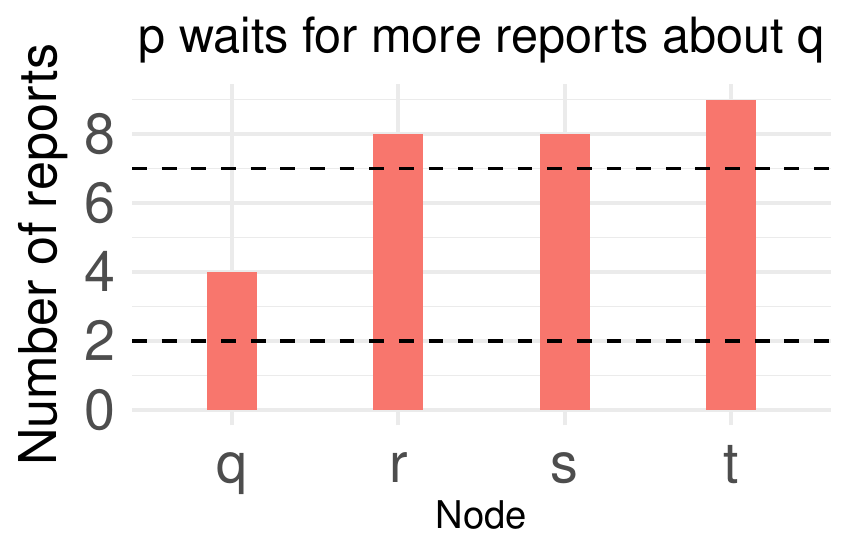}
\includegraphics[width=0.23\textwidth]{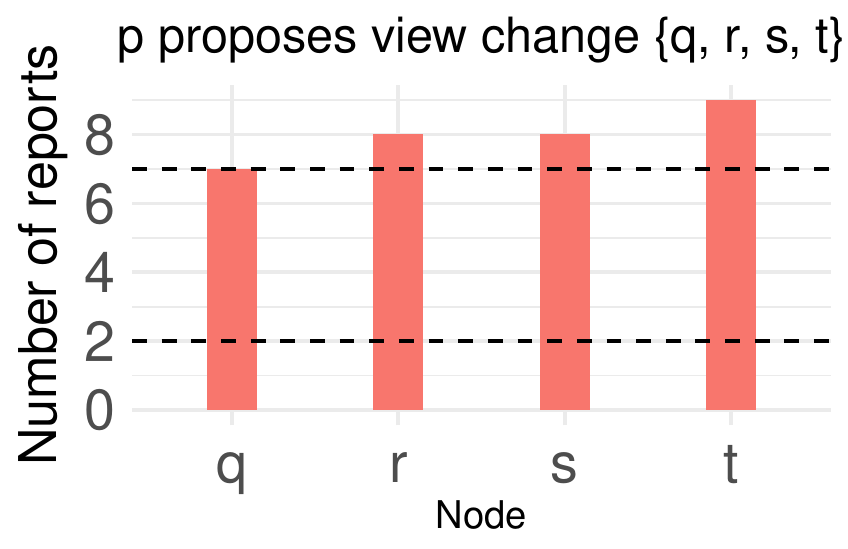}
\caption{Almost everywhere agreement protocol example at a process $p$, with
tallies about $q, r, s, t$ and $K=10,H=7,L=2$. $K$ is the number
of observers per subject. The region between $H$ and $L$ is the unstable
region. The region between $K$ and $H$ is the stable region.
\textbf{Left:} $stable=\{r, s, t\}$; $unstable=\{q\}$.
\textbf{Right:} $q$ moves from $unstable$ to $stable$; $p$ proposes
a view change $\{r, s, t\}$.}
\label{fig:watermark}
\end{figure}

Alerts in \system{} may arrive at different orders at each process. Every process
independently aggregates these alerts until a stable \CDfull{} is detected.
Our approach aims to reach agreement \emph{almost everywhere} with regards to this
detection. Our mechanism is based on a simple
key insight: A process defers a decision on a single process until the alert-count
it received on all processes is considered stable. In particular, it waits until no process'
alert-count is above a low-watermark threshold $L$ and below a high-watermark threshold $H$.

Our technique is simple to implement; it only requires
maintaining integer counters per-process and comparing them against two
thresholds. This state is reset after each configuration change.

\paragraph{Processing \remove\ and \join\ alerts} Every process ingests
broadcast alerts by \observer{}s about
edges to their \subject{}s. A \remove\ alert reports that an
edge to the \subject\ process is faulty; a \join\ alert indicates that an edge to the \subject\ is
to be created. By design, a \join\ alert can only be about a process not in the current configuration $\mathcal{C}$,
and \remove\ alerts can only be about processes in $\mathcal{C}$.
There cannot be \join\ and \remove\ alerts about the same process in $\mathcal{C}$.

Every process \emph{p} tallies up distinct \remove\mbox{} and \join\mbox{}
alerts in the current configuration view as follows. For each
\observer/\subject\ pair ($o$, $s$), \emph{p} maintains a value $M$($o$, $s$)
which is  set to $1$ if an alert was received from \observer\ $o$ regarding
\subject\ $s$;  and it is set to (default) $0$ if no alert was received.
A \emph{tally($s$)} for a process \ $s$ is the sum of entries $M$($\ast$, $s$).

\paragraph{Stable and unstable report modes} We use two parameters $H$ and
$L$, $1 \leq L \leq H \leq K$.  A process $p$ considers a process $s$ to be in
a \emph{stable report mode}   if $|\text{tally}(s)| \ge H $ at $p$.  A stable
report mode indicates that $p$ has received at least $H$ distinct \observer\
alerts about $s$, hence we consider it ``high fidelity'';  A process $s$ is
in an \emph{unstable report mode} if tally($s$) is in between $L$ and $H$. If
there are fewer than $L$ distinct \observer\ alerts about $s$, we consider
it noise. Recall that \system{} does not revert alerts; hence, a stable report
mode is permanent once it is reached. Note that, the same thresholds are used
for \remove\mbox{} and \join\mbox{} reports; this is not mandatory, and is
done for simplicity.

\paragraph{Aggregation}  Each process follows one simple rule for aggregating
tallies towards a proposed configuration change: \emph{delay proposing a
configuration change until there is at least one process in stable report mode
and there is no process in unstable report mode}.
Once this condition holds, the process announces a configuration change proposal consisting of all
processes in stable report mode, and the current configuration identifier.
The proposed configuration change has the almost-everywhere agreement property,
which we analyze in \S\ref{sec:almosteverywhere} and evaluate in \S\ref{sec:evaluation}.
Figure~\ref{fig:watermark} depicts the almost everywhere agreement mechanism
at a single process.

\paragraph{Ensuring liveness: implicit detections and reinforcements} There
are two cases in which a \subject\ process might get stuck in an unstable report
mode and not accrue  $H$ \observer\ reports.  The first is when
the \observers\ themselves are faulty.  To prevent waiting for stability
forever,  for each \observer\ $o$ of $s$,  if both $o$ and $s$ are in the
unstable report mode, then an \emph{implicit-alert} is applied from $o$ to
$s$ (i.e., an implicit \remove\  if $s$ is in $\mathcal{C}$ and a \join\
otherwise; $o$ is by definition always in $\mathcal{C}$).

The second is the rare case where the \subject\ process has good connections to
some \observers, and bad connections to others.  In this case, after a
\subject\ $s$ has been in the unstable reporting mode for a certain
timeout period,  each \observer\ $o$ of $s$ \emph{reinforces} the
detection: if $o$ did not send a \remove\ message about $s$ already, it
broadcasts a \remove\ about $s$ to echo existing \remove\mbox{}s.

\subsection{View-change Agreement}
\label{sec:almost_everywhere_agreement_to_full_agreement}

We use the result of each process' \CD{} proposal
as input to a consensus protocol that drives agreement on a single \Viewfull{}.

The consensus protocol in \system{} has a fast, leaderless path in the common case,
that has the same overhead as simple gossip. The
fast path is built around the Fast Paxos algorithm~\cite{fast_paxos}. In our
variation, we use the \CD{} result as initial input to processes, instead of
having an explicit proposer populating the processes with a proposal.
Fast Paxos reaches a decision if there is
a quorum larger than three quarters of the  membership set with an identical proposal.
Due to our prudent almost-everywhere \CD{} scheme, with high probability
, all processes indeed  have
an identical \CDfull{} proposal. In this case,
the \View{} protocol converges simply by counting
the number of identical \CD{} proposals.

The counting protocol itself uses gossip to disseminate and aggregate  a
bitmap of ``votes'' for each unique proposal. Each process sets a bit in the
bitmap of a proposal to reflect its vote. As soon as a process has a proposal
for which three quarters of the cluster has voted, it decides on that
proposal.

If there is no fast-quorum support for any proposal because there are
conflicting proposals, or a timeout is reached, Fast Paxos falls back to a
recovery path, where we use classical Paxos~\cite{Lamport:1998:PP:279227.279229}
to make progress.

In the face of partitions~\cite{Gilbert:2012:PCT:2360751.2360958}, some
applications may need to maintain availability everywhere (AP), and others
only allow the majority component to remain live to provide strong consistency
(CP).  \system{} guarantees to reconfigure processes in the majority component.
The remaining processes are forced to logically depart the system.  They may
wait to rejoin the majority component, or choose to form a separate
configuration (which \system{} facilitates quickly). The history of the members
forming a new configuration will have an explicit indication of these events,
which applications can choose to use in any manner that fits them
(including ignoring).

\section{Logically Centralized Design}
\label{sec:algorithm_aux}

We now discuss how Rapid runs as a logically centralized service, where a set
of auxiliary nodes $S$ records the membership changes for a cluster
$\mathcal{C}$. This is a similar model to how systems use ZooKeeper to manage
membership: the centralized service is the ground truth of the membership
list.

Only three minor modifications are required to the protocol discussed in \S\ref{sec:algorithm}:

\begin{enumerate}[nolistsep]
    \item Nodes in the current configuration $\mathcal{C}$ continue monitoring
    each other according to the k-ring topology (to scale the monitoring load). Instead of gossiping these
    alerts to all nodes in $\mathcal{C}$, they report it only to all nodes in $S$ instead.
    \item Nodes in $S$ apply the CD protocol as before to identify a membership change
    proposal from the incoming alerts. However, they execute the VC protocol only among themselves.
    \item Nodes in $\mathcal{C}$ learn about changes in the membership through notifications
    from $S$ (or by probing nodes in $S$ periodically).
\end{enumerate}

The resulting solution inherits the stability and agreement
properties of the decentralized protocol, but with reduced resiliency guarantees;
the resiliency of the overall system is now bound to that of $S$ 
($F = \frac{S}{2} - 1$) -- as with any logically centralized design. For progress, members of
$\mathcal{C}$ need to be connected to a majority partition of $S$. 

\section{Implementation} 
\label{sec:implementation}

\system{} is implemented in Java with 2362 lines of code (excluding comments and
blank lines).  This includes all the code associated with the membership
protocol as well as messaging and failure detection.  In addition, there are
2034 lines of code for tests. Our code is open-sourced under an Apache 2 license~\cite{rapidgithub}.

Our implementation uses gRPC and Netty for messaging. The counting step for
consensus and the gossip-based dissemination of alerts are performed over UDP.
Applications interact with \system{} using the APIs for joining and receiving
callbacks described  in \S\ref{sec:design}. The logical identifier
(\S\ref{sec:design}) for each process is  generated by the \system{} library using
UUIDs. The join method allows users to supply edge failure detectors to use.
Similar to APIs of existing systems such as Serf~\cite{serf} and Akka
Cluster~\cite{akka}, users associate application-specific metadata with the
process being initialized (e.g., \emph{"role":"backend"}).

Our default failure detector has observers send probes to their subjects and
wait for a timeout. Observers mark an edge faulty when the number of
communication exceptions they detect exceed a threshold (40\% of the last 10 measurement attempts fail). 
Similar to Memberlist and Akka Cluster,
\system{} batches multiple alerts into a single message before sending them on the
wire.

\section{Evaluation}
\label{sec:evaluation}

\paragraph{Setup} We run our experiments on a shared internal cloud service
with 200 cores and 800 GB of RAM (100 VMs).  We run multiple processes per VM,
given that the workloads are not CPU bottlenecked.  We vary the number
of processes ($N$) in the cluster from 1000 to 2000.

We compare \system{} against \emph{(i)} ZooKeeper~\cite{zookeeper} accessed
using Apache Curator~\cite{curator}, \emph{(ii)} Memberlist~\cite{Memberlist},
the SWIM implementation used by Serf~\cite{serf} and Consul~\cite{consulLink}. 
For \system{}, we use the decentralized variant unless specified
otherwise (\system{}-C, where a 3-node ensemble manages the membership
of N processes).

 We also tested Akka Cluster~\cite{akka} but found its
bootstrap process to not stabilize for clusters beyond 500 processes, and
therefore do not present further (see \S\ref{sec:motivation} and
Figure~\ref{fig:flipflopAcMlOnly}). All ZooKeeper experiments use a 3-node
ensemble, configured according to \cite{zookeeperAdmin}. For Memberlist, we use
the provided configuration for single data center settings (called
\emph{DefaultLANConfig}). \system{} uses $\{K, H, L\}=\{10, 9, 3\}$ for all
experiments and we also show a sensitivity analysis. We seek to answer:

\plparsep=2pt
\begin{compactitem}
    \item How quickly can \system{} bootstrap a cluster?
    \item How does \system{} react to different fault scenarios?
    \item How bandwidth intensive is \system{}?
    \item How sensitive is the almost-everywhere agreement property to the choice of K,H,L?
    \item Is \system{} easy to integrate with real applications?
\end{compactitem}

\paragraph{Bootstrap experiments}
\label{sub:bootstrap_experiment}

\begin{figure}[t]
\centering
\includegraphics[width=0.44\textwidth]{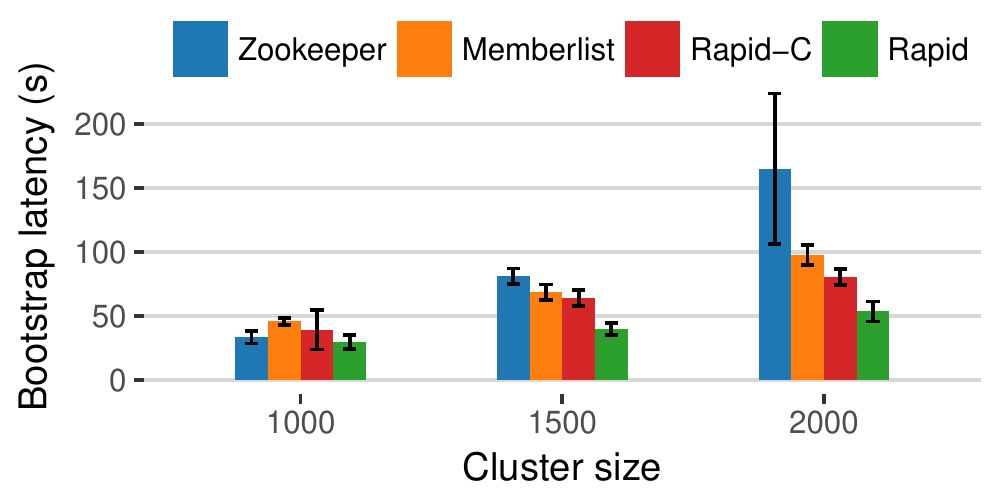}
\caption{Bootstrap convergence measurements showing the time required for all nodes to report
a cluster size of N. \system{} bootstraps a 2000 node cluster 2-2.32x faster than Memberlist, 
and 3.23-5.81x faster than ZooKeeper.}
\label{fig:bootstrap_latency}
\end{figure}

\begin{figure}[t]
\centering
\includegraphics[width=0.49\textwidth]{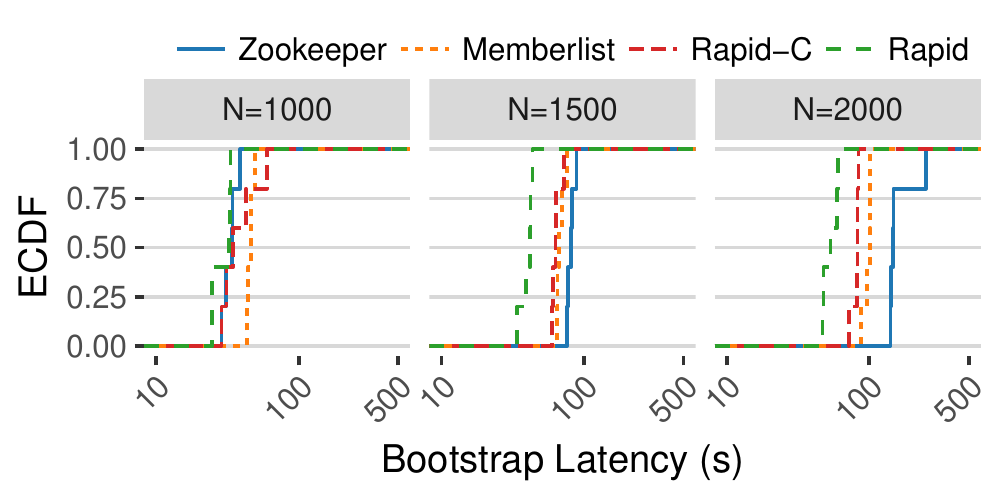}
\caption{Bootstrap latency distribution for all systems.}
\label{fig:bootstrap_latency_rapid}
\end{figure}

\begin{figure}[t]
\centering
\includegraphics[width=0.49\textwidth]{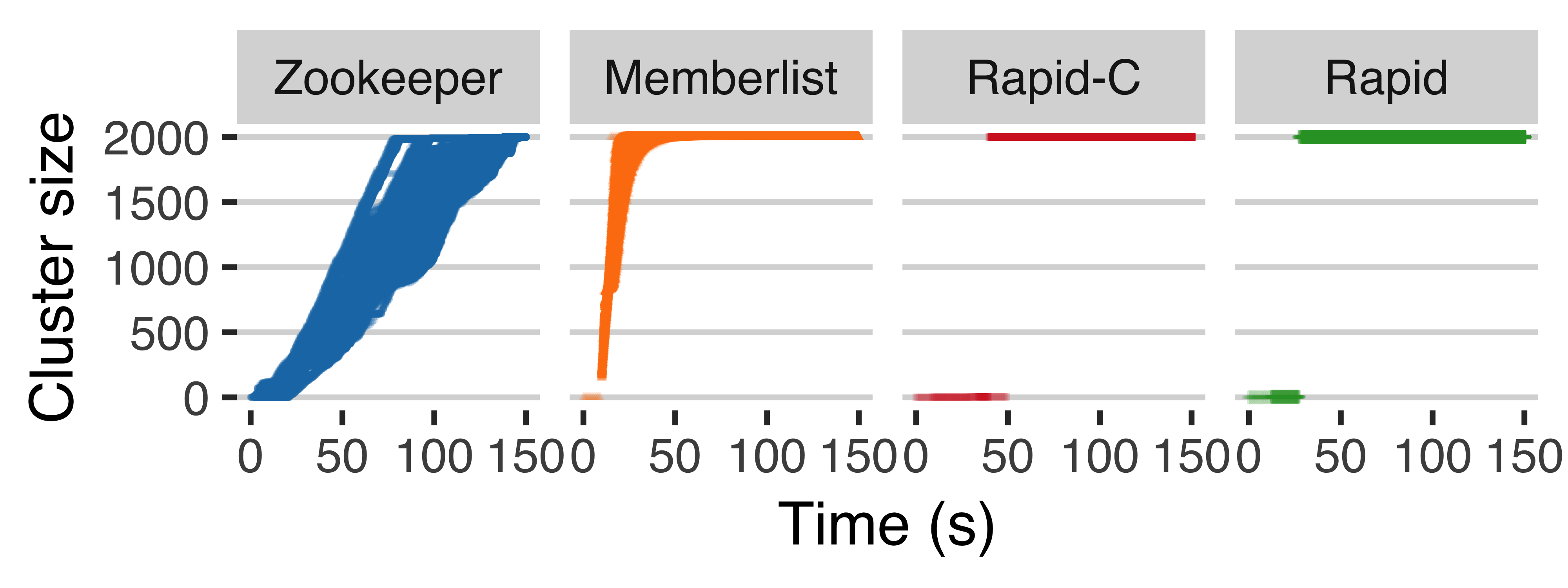}
\caption{Timeseries showing the first 150 seconds of all three systems
bootstrapping a 2000 node cluster.}
\label{fig:bootstrap_timeseries}
\end{figure}

We stress the bootstrap protocols of all three systems under varying cluster
sizes. For Memberlist and \system{}, we start each experiment with a single seed process, and after ten
seconds, spawn a subsequent group of $N - 1$ processes (for ZooKeeper, the 3-node
ZooKeeper cluster is brought up first). Every process logs its
observed cluster size every second. Every measurement is repeated
five times per value of $N$. We measure the time taken for all processes to
converge to a cluster size of $N$ (Figure~\ref{fig:bootstrap_latency}). For $N=2000$, \system{}
improves bootstrap latencies by 2-2.32x over Memberlist, and by 3.23-5.8x over ZooKeeper.

ZooKeeper suffers from herd behavior during the bootstrap process (as documented in
\cite{zookeeperHerd}), resulting in
its bootstrap latency increasing by 4x from when N=1000 to when N=2000. As we
discussed in \S\ref{sec:motivation}, group membership with ZooKeeper is done
using watches. When the $i^{th}$ process joins the system, it triggers $i - 1$
watch notifications, causing $i -1$ processes to re-read the full membership list and
register a new watch each. In the interval between a watch having triggered
and it being replaced,  a client may lose updates, leading to clients
learning different sequences of membership change events
\cite{zookeeperProgrammer}. This behavior with watches leads to the
eventually consistent client behavior in
Figure~\ref{fig:bootstrap_timeseries}. Lastly, we emphasize that this is a
3-node ZooKeeper cluster being used \emph{exclusively} to manage membership
for a single cluster. Adding even one extra watch per client to the group
node at N=2000 inflates bootstrap latencies to 400s on average.

Memberlist processes bootstrap by contacting a seed. The seed thereby
learns about every join attempt. However, non-seed processes need to periodically
execute a push-pull handshake with each other to synchronize their views (by
default, once every 30 seconds). Memberlist's convergence times are thereby as
high as 95s on average when $N=2000$ (Figure~\ref{fig:bootstrap_timeseries}).

Similar to Memberlist, \system{} processes bootstrap by contacting a seed. The
seed aggregates alerts until it bootstraps a cluster large enough to
support a Paxos quorum (minimum of three processes). The remaining processes are
admitted in a subsequent one or more view changes. For instance, in
Figure~\ref{fig:bootstrap_timeseries}, \system{} transitions from a single seed
to a five node cluster, before forming a cluster of size 2000. We confirm
this behavior across runs in Table~\ref{tab:uniquesizes}, which shows the
number of unique cluster sizes reported for different values of $N$. In the
ZooKeeper and Memberlist experiments, processes report a range of cluster sizes
between 1 and $N$ as the cluster bootstraps. \system{} however brings up large
clusters with very few intermediate view changes, reporting four and eight
unique cluster sizes for each  setting. Our logically centralized variant \system{}-C,
behaves similarly for the bootstrap process. However, processes in \system{}-C
periodically probe the 3-node ensemble for updates to the membership (the
probing interval is set to be 5 seconds, the same as with ZooKeeper).
This extra step increases bootstrap times over the decentralized variant;
in the latter case, all processes participate in the dissemination of votes through
aggregate gossip.

\begin{table}[t]
    \begin{center}
        \label{tab:uniquesizes}
        \caption{Number of unique cluster sizes reported by processes in bootstrapping experiments.}
        {\footnotesize
        \begin{tabular}{|c | c | c | c|}
            \hline
            System & N=1000 & N=1500 & N=2000 \\ \hline
            ZooKeeper & 1000 & 1500 & 2000 \\  \hline
            Memberlist & 901 & 1383 & 1858 \\ \hline
            \system{}-C & 9 & 10 & 7 \\ \hline
            \system{} & 4 & 8 & 4 \\ \hline
        \end{tabular}
        }
    \end{center}
\end{table}

\begin{figure}[t]
\centering
\includegraphics[width=0.49\textwidth]{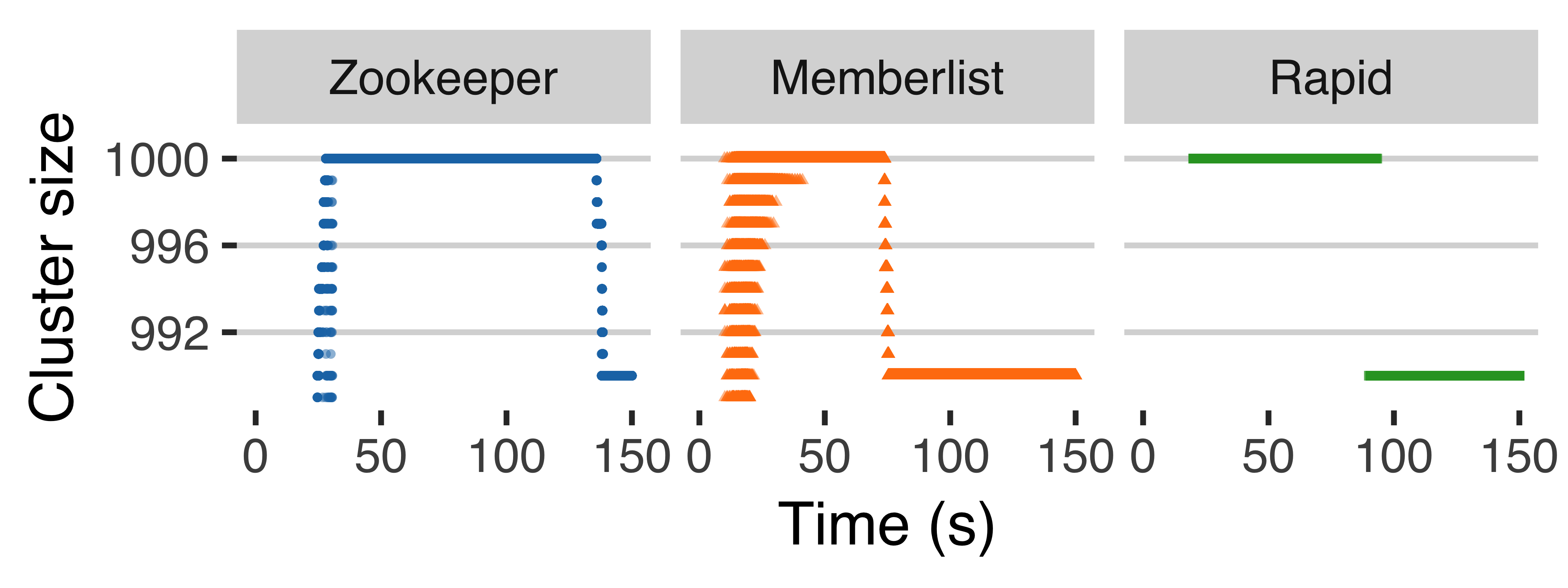}
\caption{Experiment with 10 concurrent crash failures.}
\label{fig:crash}
\end{figure}

\paragraph{Crash faults}
\label{sub:crash_faults}

We now set $N=1000$ to compare the different systems in the
face of crash faults. At this size, we have five processes per-core in the
infrastructure, leading to a stable steady state for all three systems. We
then fail ten processes and observe the cluster membership size reported by every
other process in the system.

Figure~\ref{fig:crash} shows the cluster size timeseries as recorded by each
process. Every dot in the timeseries represents a cluster size recording by a
single process. With Memberlist and ZooKeeper, processes record several
different cluster sizes when transitioning from $N$ to $N-F$. \system{} on the
other hand concurrently detects all ten process failures and removes them from
the membership using a 1-step consensus decision. Note, our edge failure
detector performs multiple measurements before announcing a fault for stability
(\S\ref{sec:implementation}), thereby reacting roughly 10 seconds later than
Memberlist does. The results are identical when the ten processes are partitioned
away completely from the cluster (we do not show the plots for brevity).

\paragraph{Asymmetric network failures}
\label{sub:flip_flop_experiments}

\begin{figure}[t]
\centering
\includegraphics[width=0.48\textwidth]{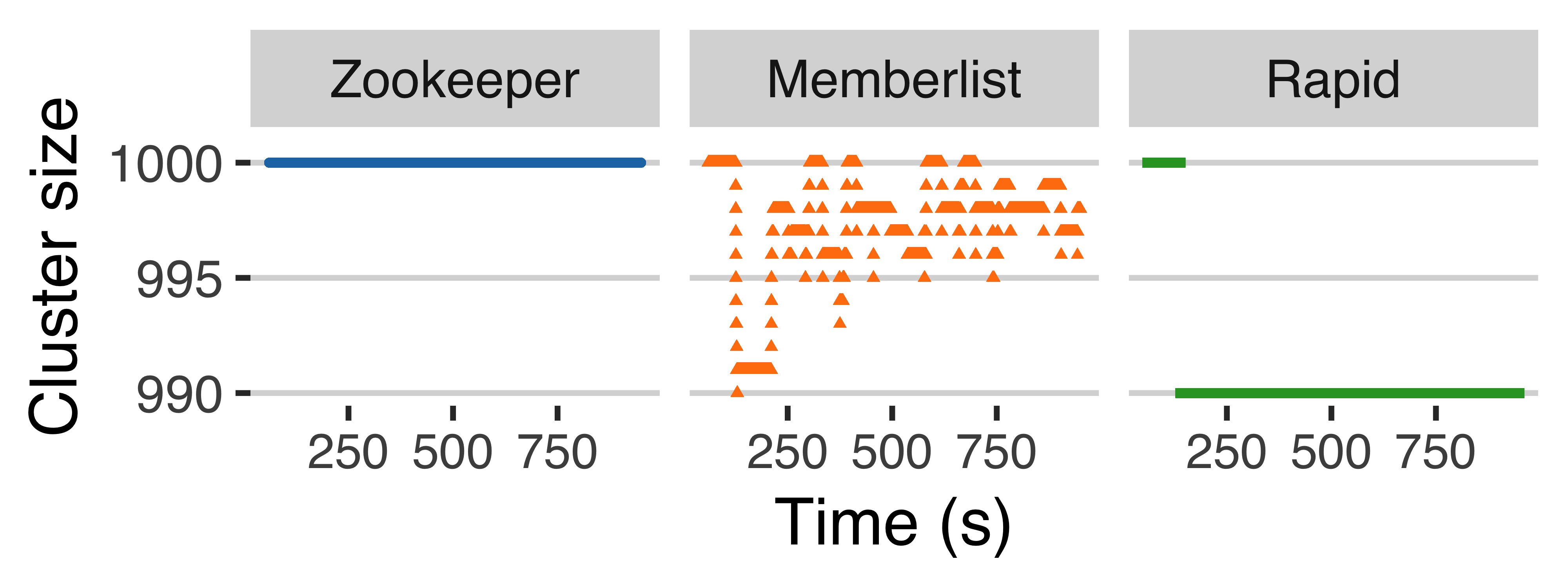}
\caption{Asymmetric network failure with one-way network partition on the network interface of
1\% of processes (ingress path). 
 }
\label{fig:flipflop}
\end{figure}

\begin{figure}[t]
\centering
\includegraphics[width=0.48\textwidth]{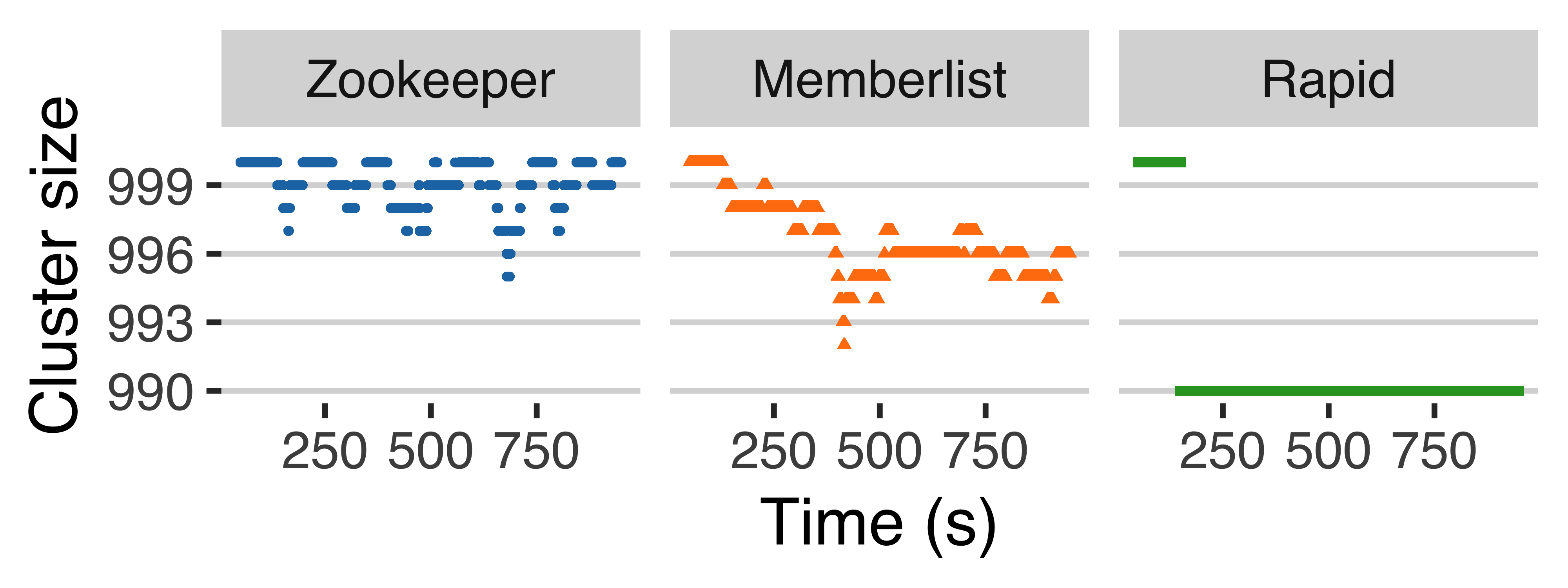}
\caption{Experiment with 80\% ingress packet loss on the network interface of 1\% of processes. 
}
\label{fig:flipflop80p}
\end{figure}

We study how each system responds to common network failures that we
have seen in practice. These scenarios have also been described in ~\cite{deanPacketLoss,
networkIsReliable, Gill:2011:UNF:2018436.2018477, turner2012failure,
Guo:2015:PLS:2785956.2787496}.

\emph{Flip-flops in one-way connectivity.} We enforce a ``flip-flopping"
asymmetric network failure.  Here, 10 processes lose all packets that they receive
for a 20 second interval, recover for 20 seconds, and then repeat the packet
dropping. We enforce this by dropping packets in the \texttt{iptables INPUT}
chain. The timeseries of cluster sizes reported by each process is shown in
Figure~\ref{fig:flipflop}.

ZooKeeper does not react to the injected failures, given that clients are not
receiving packets on the ingress path, but are able to send heartbeats to the
ZooKeeper nodes. Reversing the direction of connectivity
loss as in the next experiment does cause ZooKeeper to react.
Memberlist never removes all the faulty processes from the
membership, and oscillates throughout the duration of the failure scenario.
We also find several intervals of inconsistent views among processes. Unlike
ZooKeeper and Memberlist, \system{} detects and removes the faulty processes
(Figure~\ref{fig:flipflop}). 

\emph{High packet loss scenario.}  
We now run an experiment where 80\% of outgoing packets from the faulty processes
are dropped.  We inject the fault at $t=90s$.
Figure~\ref{fig:flipflop80p} shows the resulting membership size timeseries.
ZooKeeper reacts to the failures at $t=200s$, and does not remove all faulty
processes from the membership. Figure~\ref{fig:flipflop80p} also shows how
Memberlist's failure detector is conservative; even a scenario of sustained
high packet loss is insufficient for Memberlist to conclusively remove a set
of processes from the network.  Furthermore, we observe view inconsistencies with
Memberlist near $t = 400s$. \system{}, again, correctly identifies and removes
only the faulty processes.

\paragraph{Memory utilization.} Memberlist (written in Go) used an average of
12MB of resident memory per process. With \system{} and ZooKeeper agents (both Java
based), GC events traced using \texttt{-XX:+PrintGC} report min/max heap utilization
of 10/25MB and 3.5/16MB per process.

\paragraph{Network utilization.} Table~\ref{tab:networkUtil} shows the mean, 99th
and 100th percentiles of network utilization per second
across processes during the crash fault experiment. \system{} has a peak utilization of 9.56 KB/s received (and 11.37
KB/s transmitted) versus 7.36 KB/s received (8.04 KB/s transmitted) for
Memberlist. ZooKeeper clients have a peak ingress utilization of 38.86
KB/s per-process on average to learn the updated view of the membership.

\begin{table}[t]
\label{tab:networkUtil}
\caption{Mean, 99$^{th}$ percentile and maximum network bandwidth 
utilization per process.}
\centering
 {
 \footnotesize
     \begin{tabular}{l|ccc|}
       \cline{2-4}
       & \multicolumn{3}{|c|}{KB/s (received / transmitted)}\\
       \hline
     \multicolumn{1}{|l|}{System}     & Mean & p99 & max \\ 
       \hline
     \multicolumn{1}{|l|}{ZooKeeper}  & 0.43 / 0.01    & 17.52 / 0.33 & 38.86 / 0.67 \\ 
     \multicolumn{1}{|l|}{Memberlist} & 0.54 / 0.64    & 5.61 / 6.40 & 7.36 / 8.04  \\ 
     \multicolumn{1}{|l|}{\system{}}      & 0.71 / 0.71    & 3.66 / 3.72 & 9.56 / 11.37 \\ 
     \hline
     \end{tabular}
 }
\end{table}

\paragraph{K, H, L sensitivity study}
\label{sec:sensitivity_study}

\begin{figure}
\centering
\includegraphics[width=0.47\textwidth]{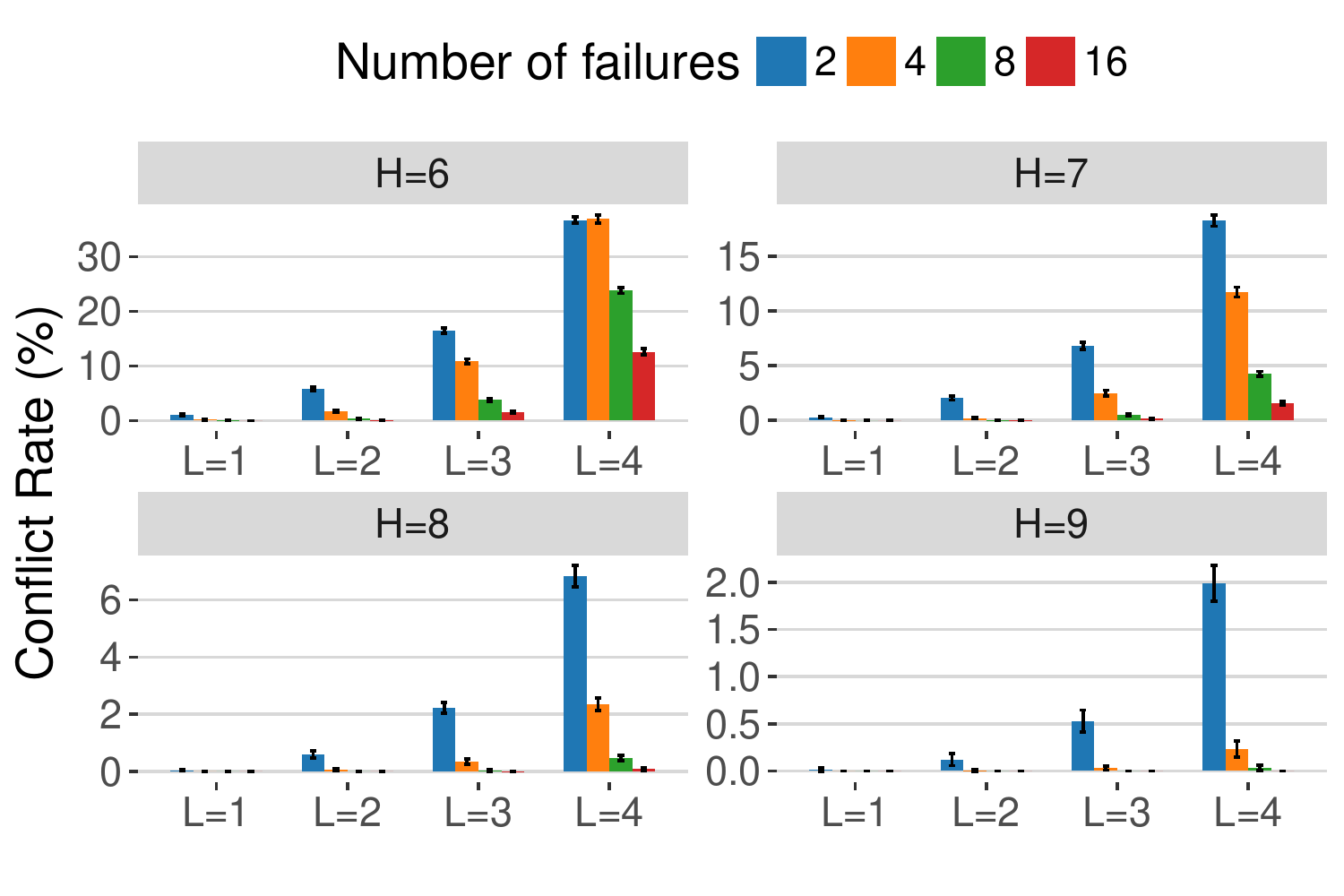}
\caption{Almost-everywhere agreement conflict probability for different combinations
of H, L and failures F when K=10. Note the different y-axis scales.
 }
\label{fig:simulation}
\end{figure}

We now present the effect of K, H and L on the almost-everywhere agreement
property of our multi-process detection technique. We initialize 1000 processes and
select $F$ random processes to fail. We generate alert messages from the $F$
processes' observers and deliver these alerts to each process in a uniform random
order. We count the number of processes that announce a membership proposal that
did not include all $F$ processes (a \emph{conflict}). We run all parameter
combinations for $H=\{6,7,8,9\}, L=\{1,2,3,4\}, F=\{2,4,8,16\}$ with 20 repetitions
 per combination.

 Figure~\ref{fig:simulation} shows the results. As our analysis
(\S\ref{sec:guarantees_summary}) predicts, the conflict rate is highest when the gap
between $H$ and $L$ is lowest ($H=6, L=4$) and the number of failures $F$ is 2.
This setting causes processes to arrive at a proposal without waiting long enough.
As we increase the gap $H-L$ and increase $F$, the algorithm at each process
waits long enough to gather all the relevant alerts, thereby diminishing the
conflict probability. Our system is thereby robust across a range of values;
for $H-L = 5$ and $F=2$, we get a 2\% conflict rate for different values of
$H$ and $L$. Increasing to $H-L=6$ drops the probability of a conflict by a
factor of 4. 

\paragraph{Experience with end-to-end workloads} 
\label{sub:end_to_end_evaluation}

We integrated \system{} within use cases at our
organization that required membership services. Our goal is to understand
the ease of integrating and using \system{}.

\begin{figure}
\centering
\includegraphics[width=0.36\textwidth]{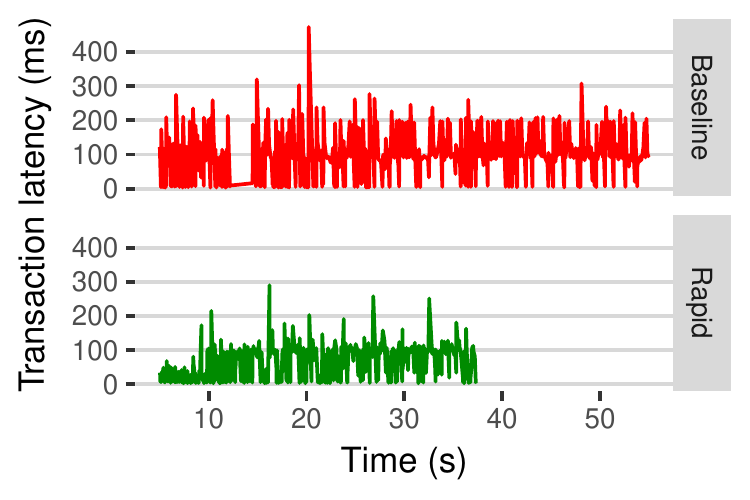}
\caption{Transaction latency when testing an in-house gossip-style failure detector and
\system{} for robustness against a communication fault between two processes. The baseline
failure detector triggers repeated failovers that reduce throughput by 32\%.}
\label{fig:corfu}
\end{figure}

\emph{Distributed transactional data platform.}  We worked with a team that uses a
distributed data platform that supports transactions.
We replaced the use of its in-house gossip-based failure detector that uses
 all-to-all monitoring, with \system{}.  The failure detector
recommends membership changes to a Paxos-based
reconfiguration mechanism, and we let \system{} provide input to the 
re-configuration management instead. Our integration added 62 and removed 25 
lines of code. We also ported the system's failure
detection logic such that it could be supplied to \system{} as an edge failure
detector, which involved an additional 123 lines of code.

We now describe a use case in the context of this system where stable
monitoring is required.  For total ordering of requests, the platform has a
transaction serialization server, similar to the one used in Google
Megastore~\cite{megastore} and Apache Omid~\cite{omid}. At any moment in time,
the system has only one active serialization server, and its failure requires the cluster to
identify a new candidate server for a failover. During this interval, workloads
are paused and clients do not make progress.

We ran an experiment where two update-heavy clients (read-write ratio of
50-50) each submit 500K read/write operations, batched as 500 transactions. We
injected a failure that drops all packets between the current serialization server and
one other data server (resembling a packet blackhole as observed by
\cite{Guo:2015:PLS:2785956.2787496}).  Note, this fault does not affect
communication links between clients and data servers. We measured the
impact of this fault on the  end-to-end latency and throughput.

With the baseline failure detector, the serialization server was
repeatedly added and removed from the membership. The repeated failovers
caused a degradation of end-to-end latency and a 32\% drop in throughput
(Figure~\ref{fig:corfu}). When using \system{} however, the system continued
serving the workload without experiencing any interruption (because no node
exceeded $L$ reports).

\begin{figure}
\centering
\includegraphics[width=0.36\textwidth]{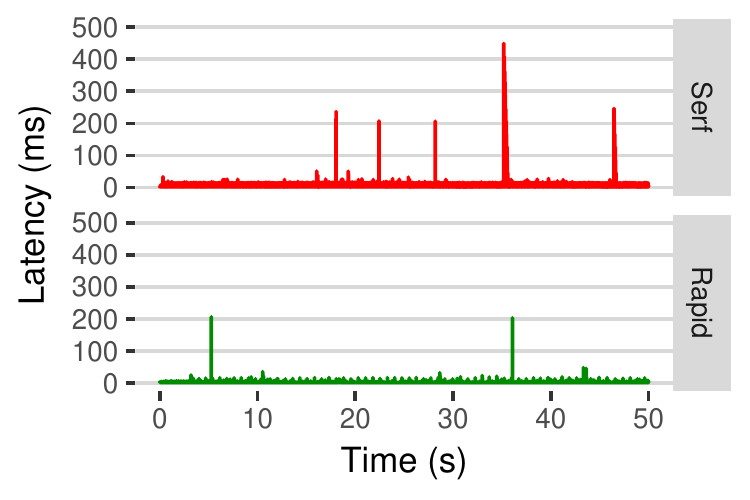}
\caption{Service discovery experiment. \system{}'s batching reduces the number of configuration reloads during a set of failures,
thereby reducing tail latency.}
\label{fig:nginxExperiment}
\end{figure}

\emph{Service discovery.} A common use case for membership is service
discovery, where a fleet of servers need to be discovered by dependent
services. We worked with a team that uses Terraform~\cite{terraform} to drive
deployments where a load balancer discovers a set of backend servers using
Serf~\cite{serf}. We replaced their use of Serf in this workflow with an agent
that uses \system{} instead (the \system{} specific code amounted to
under 20 lines of  code). The setup uses nginx~\cite{nginx} to load balance
requests to 50 web servers  (also using nginx) that serve a static HTML page.
All 51 machines run as t2.micro instances on Amazon EC2. Both membership
services update nginx's configuration file with the list of backend servers on
every membership change. We then use an HTTP workload generator to generate
1000 requests per-second.  30 seconds into the experiment, we fail ten nodes
and observe the impact on the end-to-end latency
(Figure~\ref{fig:nginxExperiment}). 

\system{} detects all failures concurrently and triggers a single reconfiguration
because of its multi-node membership change detection. Serf, which uses Memberlist,
detects multiple independent failures that
result in several updates to the nginx configuration file. The load balancer therefore
incurs higher latencies when using Serf at multiple intervals (t=35s and t=46s) because nginx is reconfiguring itself. In the steady
state where there are no failures, we observe no difference between \system{} and
Serf,  suggesting that \system{} is well suited for service discovery workloads,
despite offering stronger guarantees.

\section{Proof of correctness}
\label{sec:guarantees_summary}

Our consensus engine is standard, and borrows from the known literature on
consensus algorithms~\cite{fast_paxos,Dwork:1988:CPP:42282.42283,
Lamport:1998:PP:279227.279229}. We do not repeat its proof
of Agreement and Liveness here.

It is left to prove that faced with $F$ failures, the stable failure detector detects and
outputs $F$ at all processes with high probability. We divide the proof into
two parts.

First,  we argue in \S\ref{sec:faulty} that if $F$ consists of observably
unreachable processes, then Detection holds with high probability at all
correct processes. We consider a set $F$ consisting of faulty
processes, some of which are crashed but not necessarily observably
unreachable. We assume that $|F|$ is at most some constant fraction of
the total size say $|F| \leq n/4$. We show that the expansion property
ensures that  there exists  $T \subseteq F$ which is observably
unreachable. The set $T$ is removed from the membership view by the
consensus properties, and the system reconfigures. Then we can recurse
using the set $F` = F \setminus T$ of faulty processes. In the end, we obtain
almost everywhere Detection of all processes in $F$.
Second, in \S\ref{sec:ou} we prove the almost-everywhere agreement property about our \CDfull{} protocol.

\subsection{Detecting Faulty Processes}
\label{sec:faulty}

Let $G(V,E)$ be the (multi)-graph of all monitoring relationships,
counted with multiplicities. In other
words, $(u,v)$ is an edge if either $u$ monitors $v$ or vice versa. If
they both monitor each other, then we put two edges between them.
It follows that $G$ is $d =2K$ regular, where $K$ is the number of
rings. The graph changes dynamically as we add and remove vertices. We
will assume that $G$ is an expander throughout, meaning that its
second eigenvalue is bounded by $\lambda < d$. It is known that there
exist graphs with $\lambda \leq O(\sqrt{d})$ and indeed randomly
chosen regular graphs are known to achieve this bound with high
probability. We only assume that $\lambda/d < 1$.

We use the following standard bound on the number of edges induced within any
subset of nodes.

\begin{Lem}
  \cite[Corollary 9.2.6]{AlonSpencer}
  Let $F \subseteq V$ be an arbitrary subset of $\beta n$ vertices and let
  $e(F) $ be the number of edges in the induced subgraph of $G$ on
  $F$. Then
  \[ \left| e(F) - \frac{1}{2}\beta^2dn \right| \leq \frac{1}{2}\lambda
  \beta n.\]
\end{Lem}

A simple corollary is the following.

\begin{Cor}
  Let $S \subseteq F$ be the set of vertices in $F$ with more than $\alpha d$
  observers within $F$. Assume the following condition holds:
  \begin{align}
    \label{eq:condition}
    \beta < 2\alpha - \lambda/d.
  \end{align}
  Then $S$ is a strict subset of $F$. Indeed, we have
  \[ \frac{|S|}{|F|} \leq \frac{(\beta + \lambda/d)}{2\alpha}.\]
\end{Cor}
\begin{Proof}
Note that each element in $S$ contributes $\alpha d$ unique edges to
the set $e(F)$ of edges in the induced subgraph of $G$ on $F$. Hence\vspace{-0.1in}
\[ |S|\alpha d \leq e(F) \leq \frac{1}{2}\beta dn(\beta +
\lambda/d). \]
This implies that \vspace{-0.1in}
\[ |S| \leq  \frac{(\beta + \lambda/d)}{2\alpha}\beta n < |F|. \]

Therefore, as long as we have $\beta < 2\alpha - \lambda/d $ there will
be a non-emtpy set $T = F\setminus S$ such that every vertex in $T$
has less than $\alpha d = 2\alpha k$ observers from within $F$.
\end{Proof}

Now let us set $2\alpha = 1 - L/K$ so that Equation
\eqref{eq:condition} can be rewritten as\vspace{-0.1in}
  \begin{align}
    \label{eq:condition2}
    \beta < 1- L/K - \lambda/d.
  \end{align}
This condition says that more than $L$ neighbors of each vertex $v
\in T$ lie outside of the bad set $F$. Such vertices will be detected
as failed and removed from the cluster. Then the graph $G$ is
recomputed. We assume that it stays an expander (meaning that its
second eigenvalue is bounded by $\lambda$). We get a new bad set $F' =
F\setminus T$ in a set of $n - |T|$ vertices. The density of $F'$ is
\[ \beta' = \frac{|F| - |T|}{n -  |T|} < \frac{|F|}{n} = \beta\]
which means that Equation \eqref{eq:condition} continues to hold. This
means we can detect more failed nodes in this configuration and that
the process eventually removes all of $F$. If
we strengthen Equation \eqref{eq:condition} to say that
$\beta < 2\alpha - \lambda/d -\epsilon$ for some $\epsilon >0$, it follows that
at least a constant fraction of nodes in $F$ are removed in each round,
and so all nodes are removed in $O(\log(|F|))$ rounds.

In our experiments, with $K = 10$ (and $d =20$), we have observed
consistently that $\lambda/d < 0.45$. This means that Equation
\eqref{eq:condition2} is satisfied with $L = 3$ and $\beta = 0.25$.

\subsection{Almost-everywhere Detection}
\label{sec:ou}

\newcommand{\ac}[1]{ac_{#1}}

We show that if multiple processes fail simultaneously, the chance of
one of them exceeding the threshold of $H$ before the other reaching $L$ is
exponentially small in $K$ where the base of the exponent depends on
the separation between $H$ and $L$.

\emph{Model.} We are given a cluster with $n$ processes in which
there are $t$ ``failed" processes $a_1, \ldots, a_t$. Since the algorithm for joins
and removals is the same, we simplify the description by only referring to failures.
Each process has its own set of \observer{}s $O_i$ containing $K$ vertices each. Let $O = O_i$ denote the set of
all \observer{}s. We assume that $O_i \cap O_j = \phi$, so $|O| = tk$
(although not essential, this simplifies our analysis). The processes $a_i$ fail
simultaneously, and then each set of \observer{}s $O_i$ broadcasts \remove{}s
to the rest of the processes. Non-faulty processes receive $tK$ failure
messages with high probability. To aid the analysis, assume that $L =
\delta K, H = (1 - \delta)K$ for $0 < \delta <1/2$.

For a process $z$, we consider the bad event $B(z)$ defined as follows: {\em
There exists $S \subsetneq [t]$ where  $1 \leq s = |S| < K$ such that  $z$
receives $H$ or more messages from $O_i$ for every $i \in S$, before it
receives more than $L$ messages from any $j \in \bar{S}$.}

We will bound the probability of this event under the assumption that the set
$O$ of observers for the processes are chosen uniformly at random from all $n$
processes, and then partitioned into $\{O_i\}_{i=1}^t$. In practice, because our K-ring topology
chooses each $O_i$ uniformly at random from all $n$ processes, $O$ is the
union of all $O_i$, so there can be collisions.  But as long as $Kt
\ll \sqrt{n}$, the chances of collision are small enough that we can
ignore them, and consider the two processes identical.

\emph{Two Failures.} Let us start with $t =2$. In this case, there are two
processes $a_1, a_2$, and $B(z)$ is the event that $z$ receives $H$
messages from one process (say $a_1$) before it receives $L$ from the other.

Let us look at the set $P$ of the \observer{}s corresponding to the
first $K$ messages arriving at $z$. In
general, there are
\begin{align}
  \label{eq:denominator}
  {2k \choose k} \geq \frac{2^{2k}}{2\sqrt{k}}
\end{align}
ways of picking $P$ from $O_1 \cup O_2$ (where we use a standard lower
bound for the middle binomial coefficient).
If the event $B(z)$ happens, then $P$ contains $h \geq H$
messages from $O_1$ and $\ell = K -h$ messages from $O_2$. The
number of choices for this is
\begin{align}
  \label{eq:numerator}
  \sum_{h \geq H}{K \choose h}{K \choose K - h}
  & \leq \left(\sum_{h \geq H}{K\choose h}\right)^2\notag \leq 2^{2H_2(\delta)K}\\
\end{align}

where $H_2(\delta)$ denotes the binary entropy function, and we have
used the following approximation for the volume of the Hamming ball in $\{0,1\}^k$:
\begin{align*}
  \sum_{h \geq (1- \delta) K} {K \choose h} = \sum_{l \leq \delta K}{K \choose l} \leq 2^{H_2(\delta) K}.
\end{align*}
Plugging in the bounds from Equations \eqref{eq:denominator} and
\eqref{eq:numerator}, and accounting for the two choices for which of $a_1, a_2$ is
above $H$, we get
\begin{align*}
  \Pr[B(z)] & \leq \frac{2\cdot 2^{2H_2(\delta)K}}{2^{2K}/2\sqrt{K}} \leq
  \frac{4\sqrt{K}}{2^{2(1 - H_2(\delta))K}}\\
  & = 2^{-2(1 - H_2(\delta) - o_K(1))K}.
\end{align*}

As $\delta$ becomes smaller, the gap between $H$ and $L$ increases,
and the probability of $B(z)$ grows smaller, as one would expect. A
similar effect happens when we increase $K$.

\emph{$t$ failures.} In the general case where there are $t > 2$ processes, we can repeat the
above analysis for every pair of failed processes $a_i, a_j$ and then use
the union bound, to conclude that
\[ \Pr[B(z)] \leq {t \choose 2}2^{-2(1 - H_2(\delta) -o_K(1))K}. \]

\section{Conclusions}
\label{sec:conclusions}

In this paper, we demonstrated the effectiveness of detecting cluster-wide \CDfull{}
conditions,
as opposed to detecting individual node changes. The high fidelity of the
\CD{} output prevents frequent oscillations or incremental changes when faced
with multiple failures.
It achieves unanimous detection almost-everywhere, enabling 
a fast, leaderless consensus protocol to drive membership changes. 
Our  implementation  successfully bootstraps clusters of
2000 processes 2-5.8x times faster than existing solutions, while being stable against complex network failures.  We found \system{} easy to integrate end-to-end within 
 a distributed transactional data platform and a service
discovery use case.

{\footnotesize 
\bibliographystyle{acm}
\bibliography{refs}
}

\end{document}